\documentclass[letterpaper, 10 pt, conference]{ieeeconf}  

\IEEEoverridecommandlockouts                              

\usepackage{cite,amsmath,amssymb,amsfonts, lipsum, mathtools, cuted, hyperref,graphicx}
\let\labelindent\relax    
\usepackage{enumitem}
\usepackage{xcolor}

\usepackage{ntheorem}
\usepackage{thmtools}
\newtheorem{thm}{Theorem}

\newtheorem{lem}{Lemma}
\newtheorem{defn}{Definition}

\newtheorem{assum}{Assumption}
\newtheorem{rem}{Remark}
\newtheorem{prob}{Problem}

\allowdisplaybreaks

\title{\LARGE \bf
Data-Driven Sensor Fault Diagnosis with Proven Guarantees using Incrementally Stable Recurrent Neural Networks
    
}

\author{Farhad Ghanipoor$^{1}$, Carlos Murguia$^{1,2}$, Giancarlo Ferrari Trecate$^{3}$, and Nathan van de Wouw$^{1}$
    \thanks{$^{1}$ Mechanical Engineering Department, Eindhoven University of Technology, Eindhoven, the Netherlands. E-mails: f.ghanipoor@tue.nl, c.g.murguia@tue.nl, and n.v.d.wouw@tue.nl.}
	\thanks{$^{2}$ Engineering Systems Design, Singapore University of Technology and Design, Singapore. E-mail: {murguia\_rendon@sutd.edu.sg}.}
	\thanks{$^{3}$ Institute of Mechanical Engineering, École Polytechnique Fédérale de Lausanne (EPFL), Lausanne, Switzerland. {Email: giancarlo.ferraritrecate@epfl.ch}.}}

\begin{document}

	\maketitle
	\thispagestyle{empty}
	\pagestyle{empty}

\begin{abstract}

Robust Recurrent Neural Networks (R-RENs) are a class of neural networks that have built-in system-theoretic robustness and incremental stability properties. In this manuscript, we leverage these properties to construct a data-driven Fault Detection and Isolation (FDI) method for sensor faults with proven performance guarantees. The underlying idea behind the scheme is to construct a bank of multiple R-RENs (acting as fault isolation filters), each with different levels of sensitivity (increased or decreased) to faults at different sensors. That is, each R-REN is designed to be specifically sensitive to faults occurring in a particular sensor and robust against faults in all the others. The latter is guaranteed using the built-in incremental stability properties of R-RENs. The proposed method is unsupervised (as it does not require labeled data from faulty sensors) and data-driven (because it exploits available fault-free input-output system trajectories and does not rely on dynamic models of the system under study). Numerical simulations on a roll-plane model of a vehicle demonstrate the effectiveness and practical applicability of the proposed methodology.

		
	\end{abstract}
	
	\begin{keywords}
		Fault Detection and Isolation (FDI);
		data-driven fault diagnosis;
		recurrent neural networks;  
		convergent systems;
		incremental Integral Quadratic Constraints (IQC).
	\end{keywords}

	\section{INTRODUCTION} \label{sec: introduction}
	
The increasing demand for high reliability in modern engineered systems, driven by the need for product quality and cost efficiency, has made fault detection and isolation (FDI) crucial. Unexpected failures, particularly in safety-critical and high-precision applications, can cause severe operational disruptions and substantial economic losses. To mitigate these risks, advanced FDI methods are essential for early fault detection, preventing catastrophic failures, lowering maintenance costs, and minimizing unplanned downtime. By leveraging FDI techniques, engineered systems can operate with greater efficiency and safety, ensuring uninterrupted production and long-term sustainability \cite{isermann2006fault},\cite{ding2008model}.

Model-based FDI methods have been an active area of research over the past two decades \cite{chen2012robust,steven2013model,zhu2015fault,pertew2007lmi, ghanipoor2023linear,ghanipoor2025robust}. These methods rely on accurate system models to characterize deviating behavior of measured system data from the behaviour of such models. In such appraoaches, first, system models are identified using input-output data and, second, model-based fault diagnosis methods are designed \cite{ossmann2016enhanced, simani2003model, hwang2009survey}. While model-based methods are reliable and provide performance guarantees, in practice, obtaining accurate system models is challenging, time-consuming, and often impossible due to the high complexity of the system under study. 

Therefore, recent research has explored data-driven methods that bypass the need for explicit system identification, instead leveraging the foundational connection between system identification and model-based fault diagnosis techniques \cite{ding2014data, wan2016data}. These data-driven approaches simplify FDI schemes by omitting the explicit system identification step while still achieving comparable performance to model-based methods. Hereafter, we refer to such data-driven techniques as \emph{direct methods}.


The development of direct methods is still ongoing, even for linear systems. For instance, \cite{sheikhi2024kernel} and \cite{edelmayer2007inversion} propose system inversion-based methods, which reconstruct unknown inputs or disturbances by reversing system dynamics — an approach that enables fault estimation without requiring an explicit system model. These inversion-based methods have also been extended to nonlinear systems \cite{edelmayer2004input}. Another class of direct methods for nonlinear systems are projection-based schemes, which map the system dynamics into a higher-dimensional space to establish a linear representation \cite{wang2016kernel, si2020key}. This allows the use of linear data-driven techniques for the design of  fault detection and isolation (FDI) schemes.

Given the flexibility of Deep Neural Networks (DNNs) in learning complex high-dimensional functions, they have been used to learn direct mappings from input-output data to unknown fault signatures \cite{ding2008model,shahnazari2020fault, talebi2008recurrent}. While neural network-based methods demonstrate high accuracy, they often lack theoretical performance guarantees for fault diagnosis. The latter observation motivates the development of DNNs-based FDI methods for nonlinear systems with provable diagnostic guarantees that enable the isolation of single, multiple, and simultaneous additive faults. In this paper, we will develop such a method for sensor faults. 

Leveraging recent results in \cite{revay2023recurrent}, we exploit Robust Recurrent Neural Networks (R-RENs), a class of recurrent neural networks with a direct parameterization that inherently guarantees the satisfaction of input-output incremental IQCs (Integral Quadratic Constraints), without requiring additional constraints during training. The use of incremental IQC properties provides provable diagnostic guarantees. This property, when applied to the construction of a bank of fault detectors, ensures that fault detector filters remain insensitive to faults for which they are not specifically designed. Furthermore, the direct parameterization of R-RENs simplifies learning, enabling the use of standard optimization techniques such as stochastic gradient descent. Additionally, R-RENs are highly expressive, comprising stable linear systems, contracting recurrent neural networks, deep feed-forward networks, and others \cite{revay2023recurrent}.

For sensor fault diagnosis, we propose to exploit a bank of R-RENs (acting as fault isolation filters), each designed to be sensitive to faults occurring in a particular sensor and robust against faults in all the others. The proposed method is unsupervised (as it does not require labeled data from faulty sensors) and fully data-driven (because it exploits available fault-free input-output system trajectories and does not rely on dynamic models of the system under study). Then, if faults occur in some sensors, their corresponding R-RENs will return increased fault signatures in response to measured input-output trajectories, which will enable fault isolation. 

The main contributions of this paper are as follows: 
	\begin{enumerate} [label=(\alph*)]
		\item 	 \textbf{{Model-Free FDI Design}}: Using only healthy input-output data, we propose a direct method for sensor fault detection and isolation for discrete-time nonlinear systems that can address single, multiple, and simultaneous sensor fault occurrences.   
		\item \textbf{{Performance Guarantees}}: We prove that each R-REN-based filter is globally exponentially convergent for bounded inputs, utilizing the contractivity property of R-RENs. Moreover, given the build-in satisfaction of input-output incremental IQCs for the R-RENs, we guarantee that fault detector filters remain insensitive to faults for which they are not specifically designed. Additionally, we establish a necessary condition to ensure fault sensitivity for each filter with respect to the associated designed fault. 
    	\end{enumerate}
	
The content of the remainder of the paper is as follows. The problem formulation for sensor fault detection and isolation for nonlinear discrete-time systems is described in Section~\ref{sec:problem_formulation}. The proposed solution is outlined in Section~\ref{sec:problem_solution}. The results of this approach are evaluated via numerical results in Section~\ref{sec:numrical_results}. The final remarks, together with recommendations for future work, are provided in Section~\ref{sec:conclusion}.
	
		\textbf{Notation:} The identity matrix of size $n \times n$ is denoted as $I_n$ or simply $I$ when the context specifies $n$. Similarly, matrices of dimensions $n \times m$ comprising only zeros are denoted as ${0}_{n \times m}$ or ${0}$ when the dimensions are clear. The set of sequences $x: \mathbb{N} \rightarrow \mathbb{R}^n$ is denoted by $\ell_{2 e}^n$.  Superscript $n$ is omitted when it is clear from the context. For $x \in \ell_{2 e}^n,$ $x(k)$ is the value of the sequence $x$ at sample $k \in \mathbb{N}$. For notation simplicity, we often use $x$ and $x^+$ instead of $x(k)$ and $x(k+1)$, respectively. The subset $\ell_{2}\subset \ell_{2 e}^n$ consists of all square-summable sequences, i.e., $x \in \ell_{2}$ if and only if the $\ell_{2}$ norm $\|x\|:= \sqrt{\sum_{k=0}^{\infty}\left|x(k)\right|^2}$ is finite, where $|\cdot|$ denotes Euclidean norm. Given a sequence $x \in \ell_{2 e}^n$, the $\ell_{2}$ norm of its truncation over $[0,k_f]$ is $\|x\|_{k_f}:=\sqrt{\sum_{k=0}^{k_f}\left|x(k)\right|^2}$. A positive definite matrix is symbolized by $P \succ 0$. For a positive definite symmetric matrix $P$ and a vector $x$, $|x|_P$ is the weighted norm $\sqrt{x^\top P x}$. For column vectors $x$ and $y$, we write $\text{col}[x,y] := [x^\top,y^\top]^\top$ 
	
	\section{Problem Formulation} \label{sec:problem_formulation}
	Consider the discrete-time nonlinear system:
	\begin{equation}   \label{eq:sys}
		\left\{\begin{aligned}
			x^+ &= g(x, u),\\
			y &= h(x,u) +  f,
		\end{aligned}\right. 
	\end{equation}
	where $x \in {\mathbb{R}^{n}}$, $y \in {\mathbb{R}^{{m}}}$, and $u \in {\mathbb{R}^{{l}}}$ are the system state, measured output and known input, respectively, with ${n},{m}, l \in \mathbb{N}$. Functions $g: \mathbb{R}^{n} \times \mathbb{R}^{l} \to \mathbb{R}^{n}$ and $h: \mathbb{R}^{n} \times \mathbb{R}^{l} \to \mathbb{R}^{m}$ describe the system dynamics and output mapping, respectively. Signal $f: \mathbb{Z}  \to \mathbb{R}^{m}$ denotes the unknown sensor fault vector. We assume no model for system \eqref{eq:sys} is available. We seek to design a scheme to detect and isolate fault occurrences in individual or multiple entries of the measured output (i.e., faults in one or more sensors), given input-output ($u,y$) data only.
	
Inspired by model-based FDI methods, we propose an isolation scheme based on a bank of filters. Each filter aims to detect a fault in a particular sensor while robustifying the impact of faults in all the others. The $i$-th filter generates a residual $r_i$, $i = 1,\ldots,m$, which serves as an indicator for fault occurrence in the $i$-th sensor. These residuals are then used by a decision-making algorithm to determine the system status (healthy/faulty) and identify the faulty sensor (if the system is faulty). Here, the focus of this work is on the design of the filter bank, not the decision-making process.

	
Without loss of generality, the structure of each filter can be characterized as follows: 
	\begin{equation} \label{eq:filter_structure}
		\left\{\begin{aligned}
			z^+_i &=  a_i (z_i,\bar{u}; \theta_i ), \quad i =1,\ldots,m,\\
			r_i&=  b_i (z_i,\bar{u}; \theta_i),
		\end{aligned}\right. 
	\end{equation}
where $z_i \in {\mathbb{R}^{n_z}}$ represents the internal filter state, and the scalar residual $r_i$ is the output of the filter. We collect inputs and outputs in the stacked vector $\bar{u} :=\text{col}[u,y] \in \mathcal{\bar U}$, for some compact set $\mathcal{\bar U} \subset \mathbb{R}^{l+m}$. The filter functions $a_i(\cdot)$ and $ b_i(\cdot)$, and parameters $\theta_i,  i =1,\ldots,m$, are to be designed. Here, we aim to design well-posed filters that converge (independent of their initial condition) to a unique, asymptotically stable steady-state solution, associated to the bounded filter input $\bar{u} \in \mathcal{\bar U}$. Such property guarantees that despite the nonlinearity of the filter, each measured input-output pair $(\in \bar{u})$ will lead (after a transient) to a unique residual signature in the output of the filter. These properties are effectively represented by the concept of convergence (see \cite{jungers2024discrete, tran2018convergence, pavlov2011steady}). In the following definition, we introduce the notion of convergence for the FD filters in \eqref{eq:filter_structure}.
	
	\begin{defn}\emph{\textbf{(Convergence)}} 
		The fault detector filters \eqref{eq:filter_structure} are said to be globally exponentially convergent for a class of bounded inputs $\mathcal{\bar U}$, if, for every $\bar u \in \mathcal{\bar U}$, and for all $i =1,\ldots,m$:
		\begin{itemize}
			\item there exists a unique and bounded steady-state solution $\bar{z}_i(k, \bar u)$ to \eqref{eq:filter_structure};\vspace{1mm}
			\item $\bar{z}_i(k, \bar u)$ is globally exponentially stable.
		\end{itemize}
	\end{defn}
	
Given the fact that no model of system \eqref{eq:sys} is available to design the filters, we pursue a data-driven approach for which a dataset, of length $s_f \in \mathbb{N}$, of fault-free input and output trajectories, $\bar{u}(k) =\text{col}[u(k),y(k)] \in \mathcal{\bar U}$, $k\in\{1,\ldots,s_f\}$, is available. We will design the filters' structure to achieve the desired filter properties (convergence) and propose a training method to extract fault signatures. Because we do assume that no labeled data of faulty sensors is available, we need a mechanism to extract their signatures given healthy data $\bar{u}$. To this end, for every sensor measurement $y_i(k)$, we create \textit{synthetic data of sensor faults}, $f_i^s(k)$, and use it to create $m$ synthetic datasets $d_i := \{u(k),y_i(k)+ f_i^s(k),y_j(k)\}$, $i,j \in \{1,\ldots,m\}$, $j \neq i$, $k\in\{1,\ldots,s_f\}$. Then, we train the $i$-th filter with input $\{u(k),y_i(k)+ f_i^s(k),y_j(k)\}$ and enforce that the $i$-th filter output $r_i$ tracks the synthetic fault signal $f_i^s(k)$. That is, during training, we minimize a cost of the form $(r_i(k)-f_i^s(k))^2$ over the data set $d_i$ and the filter structure. This will enforce that the filter predicts additive signals to the sensor measurement $y_i$. If there is no fault, the filter must return a signal close to zero, and if there is, the filter will return an estimate of the fault-induced signal. Note that the generation of synthetic faulty data incurs no cost in terms of data collection from the actual system; hence, different types of faulty data can be generated to cover various fault classes. Any prior knowledge of fault types in the actual system can assist in selecting appropriate synthetic faulty data. Conversely, if no prior is known, we could, e.g., train on synthetic Fourier series with a large number of harmonics to increase the expressivity of the filter.

We can now state the problem we seek to solve.

	\begin{prob} \textbf{\emph{(Direct Design of FD Filters with Proven Guarantees)}} 		\label{prob:main_prob}
		Consider the data-generating system \eqref{eq:sys}, the healthy dataset $\bar{u}(k)$, the synthetic dataset of sensors faults $f_i^s(k)$, and the filters \eqref{eq:filter_structure}, $i = \{1,\ldots,m\}$, $k\in\{1,\ldots,s_f\}$. Design the filter functions $a_i(\cdot), b_i(\cdot)$, and parameters $\theta_i$ so that: \\
		\textbf{\emph{1) Optimality:}} A suitable cost function is minimized over the dataset and filter structure to enforce that the filter outputs $r_i$ track the synthetic fault signals $f_i^s$\emph{;}\\
		\emph{\textbf{2) Convergence:}} The fault detector filters \eqref{eq:filter_structure} are globally exponentially convergent for $\bar u \in \mathcal{\bar U}$\emph{;}\\
		\emph{\textbf{3) $i$-th Fault Sensitivity:}} The residual $r_i, i = 1, 2, \ldots, m,$ is sensitive to fault signals in the $i$-th sensor\emph{;}\\
		\emph{\textbf{4) $j$-th Fault Insensitivity:}} The residual $r_i$ is insensitive to fault signals in the $j$-th sensor, $j \in\{1,2, \ldots, m\}$, $j \neq i$\emph{.}
	\end{prob}
    

	
In the following section, we propose a solution to Problem~1 using robust recurrent equilibrium networks (R-RENs).
	
\section{Robust REN-Based Fault Detector Filters} \label{sec:problem_solution}		
	
We pursue a deep learning approach to the design of the fault detection filters. The design process involves solving a training problem that minimizes a cost function evaluated over the synthetic dataset (fault-free input-output trajectories and synthetic fault signals) and the filter structure.

Recent results in \cite{revay2023recurrent} introduce the notion of Robust acyclic Recurrent Equilibrium Networks (R-aRENs). Using such Robust acyclic RENs (R-aRENs) and by the grace of their contractivity we guarantee the first goal (i.e., convergence of each filter) in Problem~\ref{prob:main_prob}. Moreover, using the built-in satisfaction of input-output incremental IQCs for the R-aRENs, we address the third and fourth properties we aim to achieve in Problem~\ref{prob:main_prob}. More specifically, the incremental IQC properties of the R-aRENs, with a specific selection of design parameters (provided later), can exhibit an entry-based incremental $\ell^2$-gain from a filter's input to its output. Using this entry-based gain, for each fault detector filter, we can tune a high gain for the fault entry for which the filter is designed, leading to a necessary condition for fault sensitivity (i.e., the third goal in Problem~\ref{prob:main_prob}). Using the same gain, we can tune a low gain for the other entries (other than the fault entry), leading to a sufficient condition for insensitivity to other faults (i.e., the last goal in Problem~\ref{prob:main_prob}).

Furthermore, note that the properties of R-aRENs are \emph{built-in} behavioral guarantees. That is, to ensure the IQC properties, no parametric constraints needs to be imposed during the learning of R-aRENs. This simplifies the learning process because it allows the use of general methods for unconstrained optimization, such as, e.g., stochastic gradient descent and its variations \cite{revay2023recurrent}. Furthermore, as discussed in the introduction, RENs demonstrate high flexibility by covering different types of nonlinear systems \cite{revay2023recurrent}.
	
	Considering the general structure of the FD filters in \eqref{eq:filter_structure}, we select $a_i(\cdot)$ and $b_i(\cdot)$ such that the FD filters are in the form of R-aRENs as provided in \cite{revay2023recurrent}. Therefore, the FD filters can be written as 
	\begin{equation} \label{eq:REN_filter}
		\begin{aligned}
			\left[\begin{array}{c}
				z_i^+ \\
				v_i \\
				r_i
			\end{array}\right] &=W_i\left[\begin{array}{c}
				z_i \\
				w_i \\
				\bar u
			\end{array}\right]+\eta_i, \quad i =1,\ldots,m,\\
			w_i &=\sigma_i\left(v_i\right),
		\end{aligned}
	\end{equation}
	where $v_i, w_i \in \mathbb{R}^{n_v}$ are the input and output of activation function $\sigma_i(\cdot)$, respectively. The learnable parameters are $\theta_i :=\{W_i, \eta_i\}$ with $W_i \in \mathbb{R}^{(n_z+n_v+1) \times(n_z+n_v+(l+m))}$ as the weight matrix and $\eta_i \in \mathbb{R}^{n_z+n_v+1}$ the bias vector. The function $\sigma_i(\cdot)$ is an entry-wise function satisfying the following assumption. 
	
	\begin{assum}\emph{\textbf{(Slope-Restricted Activation Function)}} 	\label{assum:slope_restricted_nl}
		The activation functions $\sigma_i(v_{i}):=\left[\begin{array}{llll} \sigma_{i,f}\left(v_{i,1}\right) & \sigma_{i,f}\left(v_{i,2}\right) & \cdots & \sigma_{i,f}\left(v_{i,n_v}\right) \end{array}\right]^{\top},  \; i =1,\ldots,m$, is piecewise differentiable and slope-restricted in $[0,1]$, i.e.,
		$$
		0 \leq \frac{\sigma_{i,f}(y)-\sigma_{i,f}(x)}{y-x} \leq 1, \; i =1,\ldots,m, \forall x, y \in \mathbb{R}, x \neq y,
		$$
		and $\sigma_{i,f}(0) = 0$, for $ i =1,\ldots,m$. 
	\end{assum}
	
	In \cite[Thm. 1]{revay2023recurrent}, an Linear Matrix Inequality (LMI) condition is provided to ensure contractivity and incremental Integral Quadratic Constraints (IQC) to induce a robustness property for R-aRENs. This LMI condition can also be used to satisfy some of the requirements in Problem~\ref{prob:main_prob}. This means that during the learning process of FD filters, a semi-definite program with the LMI condition should be solved to ensure the aforementioned properties. Imposing the LMI constrains the R-aREN parameter $\theta_i$ to a convex set $\Theta_i \subseteq \mathbb{R}^N$, $N = (n_z+n_v+1) \times(n_z+n_v+(l+m)+1)$.
	
	Subsequently, in \cite[Sec. V]{revay2023recurrent}, a direct parameterization approach, i.e., smooth mappings from $\mathbb{R}^{N}$ to the weights and biases of an R-aREN, is proposed such that the above discussed LMI conditions are automatically sastisfied. Consequently, the R-aREN parameter $\theta_i$ is no longer restricted to belong to a convex set $\Theta_i$, enabling the application of unconstrained optimization methods for learning. In the direct parameterization in \cite{revay2023recurrent}, some elements of the weights of an R-aREN are free variables, meaning that they have no effect on the contractivity and incremental IQC properties of RENs and can be freely parameterized. On the other hand, some other elements have to be selected such that the LMI condition is automatically satisfied. We call this new set of learnable parameters (i.e., free parameters) $\theta_{d_i}$. Here, we use the direct parameterization of R-aRENs as FD filters. 
	
	In the direct parameterization of the FD filters in \eqref{eq:REN_filter}, we have the following design parameters:
	\begin{equation} \label{eq:RQ}
		\begin{aligned}
		R_i &:= \text{diag}(\beta_i I_l, \gamma_i I_{i-1}, \beta_i, \gamma_i I_{m-i}), \\
        Q_i &:= -q_i, \quad i =1,\ldots,m,
		\end{aligned}
	\end{equation}
	with positive integers $\gamma_i, \beta_i$ and $q_i$ as tunable parameters affecting fault sensitivity and other faults insensitivity of the $i$-th FD filter, see Problem~\ref{prob:main_prob}. Note that, as discussed above, there are more free parameters than those introduced here. For the sake of readability, details on the direct parameterization of the R-aRENs in \eqref{eq:REN_filter} are provided in Appendix~\ref{ap:direct_parameteriznatio}. However, the free parameters $\gamma_i, \beta_i$ and $q_i$ are explicitly introduced here because they characterize the sensitivity properties of the filters to faults (related to the requirements in Problem~1).
	
	The cost function for training of R-aRENs FD filters are given by
	\begin{equation} \label{eq:cost}
		J = \min \limits_{\theta_{d_i}} \sum_{s=0}^{s_f} \sum_{k=k_0}^{k_f} (r_{i}^s(k)- f_{i}^{s}(k))^2, \quad i = 1,2,\ldots,m, \\
	\end{equation}
	where integer $k_f$ represents the final index of samples in each data-set. Integer $k_0$ is an adjustable parameter for the initial sample. Scalars $r_{i}^s(k)$ and $f_{i}^{s}(k)$ denote the residuals of the $i$-th filter and the corresponding fault realization (of the $i$-th sensor) for scenario $s$, respectively. The following theorem formalizes the main result of this paper.
	
	\begin{thm}\textbf{\emph{(Direct FDI Filter Design with Performance Guarantees)}} 			\label{theorem}
		Consider the data-generating system \eqref{eq:sys}, $s_f$ data-sets of input-output signals, $u$ and $y$, and $s_f$ synthetic data-sets of sensors faults. Moreover, consider the R-aREN fault detection filters \eqref{eq:REN_filter} with given design parameters $R_i = \text{diag}(\beta_i I_l, \gamma_i I_{i-1}, \beta_i, \gamma_i I_{m-i}) \succ 0$, $Q_i = -q_i < 0, i = 1,2,\ldots,m$, as introduced in \eqref{eq:RQ}, and adopt Assumption~\ref{assum:slope_restricted_nl}. Additionally, impose the direct parameterization in Appendix~\ref{ap:direct_parameteriznatio} on the filters \eqref{eq:REN_filter} and train those filters using unconstrained minimization of the cost in \eqref{eq:cost}. Let $\theta_{d_i}^\star, i = 1,2,\ldots,m$, denote the optimizers. Then, setting the parameters of the FD filters \eqref{eq:REN_filter} as $\theta_i= \theta_i^\star,i = 1,2,\ldots,m$, where $\theta_i^\star$ are the parameters mapped from $\theta_{d_i}^\star$, guarantees the following properties to be guaranteed:\\
		\emph{\textbf{1) Convergence:}} The fault detector filters in \eqref{eq:REN_filter} are globally exponentially convergent for the class of inputs $\bar u \in \mathcal{\bar U}$\emph{;}\\
		\emph{\textbf{2) Fault Sensitivity:}} It holds that 
		\begin{equation} \label{eq:fault_sensitivity}
			\| \delta r_i^{s_{12}} \|_{k_f} \leq   \sqrt{\frac{\beta_i}{q_i}} \|f_i\|_{k_f}, \quad \forall k_f, i = 1,2,\ldots,m,
		\end{equation}
		where $\delta{r_i}^{s_{12}} :=  r_i^{s_2} - r_i^{s_1},$ with $r_i^{s_2}$ representing the residual of the $i$-th filter for any scenario of fault occurrence in the $i$-th entry sensor and $r_i^{s_1}$ representing any healthy scenario (i.e., $f$ zero in \eqref{eq:sys}). The scalar $f_i$ indicates the $i$-th entry of the fault vector $f$ in \eqref{eq:sys}\emph{;}\\
		\emph{\textbf{3) Other Faults Insensitivity:}} It holds that
		\begin{equation} \label{eq:faults_insensitivity}
			\|\delta r_i^{s_{13}} \|_{k_f} \leq   \sqrt{\frac{\gamma_i}{q_i}} \|\tilde{f}\|_{k_f}, \quad \forall k_f, i = 1,2,\ldots,m,
		\end{equation}
		where $ \delta{r_i}^{s_{13}} :=  r_i^{s_3} - r_i^{s_1},$ with $r_i^{s_3}$ representing the residual of the $i$-th filter for any fault occurrence scenario in sensors, except for the $i$-th entry sensor. The fault vector $f = \tilde{f}$ in \eqref{eq:sys} is a vector with at least one non-zero $j$-th entry ($j \neq i$) and zero $i$-th entry\emph{.}
	\end{thm}
	\emph{\textbf{Proof}:} 
	The proof can be found in Appendix \ref{ap:thm1_proof}.
	\hfill $\blacksquare$
	
	\begin{rem}\textbf{\emph{(Design Parameters Selection)}} 	
		Based on the results provided in Theorem \ref{theorem}, and with Problem \ref{prob:main_prob} in mind, the design parameters of the FD filters in \eqref{eq:RQ} should be selected as follows:
		$$
		0 < \gamma_i \ll q_i \ll \beta_i, \quad i = 1,2,\ldots,m.
		$$
		Specifically, based on \eqref{eq:fault_sensitivity}, $q_i \ll \beta_i$ is a necessary condition for fault sensitivity and, based on \eqref{eq:faults_insensitivity}, $\gamma_i \ll q_i$ is a sufficient condition for other faults insensitivity. Note that to satisfy the fault sensitivity and insensitivity properties in Problem~1 (i.e., the third and fourth properties), the cost function in \eqref{eq:cost} can enforce these conditions during training, as it incorporates faulty data.
	\end{rem}
	
    \section{Illustrative Example} \label{sec:numrical_results}
	
	In this section, we evaluate the proposed method through a nonlinear case study. The case study is a four-degree-of-freedom roll plane maneuver of a vehicle with nonlinear suspension \cite{kessels2023real, ghanipoor2024learning}. A schematic representation of this system is provided in Figure~\ref{fig:rpm_schematic}. The equations of motion and system parameters are detailed in Appendix~\ref{ap:rpm_app}. The system's nonlinearity arises from the nonlinear stiffness and damping between the roll bar and both tires. The available measured outputs include the relative displacement and relative velocity between the roll bar and both tires. The system inputs correspond to the road surface positions ($u_1$ and $u_2$ in Figure~\ref{fig:rpm_schematic}). 
	
	\begin{figure}[t!]
		\centering
		\smallskip
		\includegraphics[width=1\linewidth,keepaspectratio]{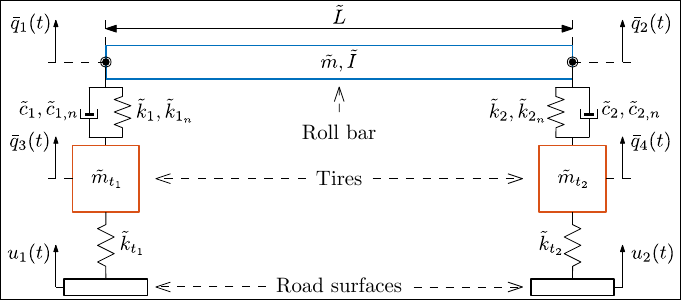}
		\caption{Roll plane system schematic.}
		\label{fig:rpm_schematic}
	\end{figure}
	
	
	For the aforementioned system, all four sensors may become faulty, and the objective is to design a bank of four filters using the proposed REN-based approach in Section~\ref{sec:problem_solution}. We assume that input-output data from the system are available, categorized into training and test data-sets, which are generated by applying different system inputs, as outlined in Section~\ref{sec:data_sets}. Synthetic fault data are generated and added to the sensor measurements. Notably, even in experimental settings, no actual faulty data are required, as they are synthesized. The process of generating synthetic fault data is also discussed in Section~\ref{sec:data_sets}. Following the proposed method, each R-aREN fault detector filter is trained using the optimization problem in \eqref{eq:cost}. The training problem \eqref{eq:cost} is implemented in PyTorch, and the R-aRENs (which can be seen as deep neural networks) are trained using ADAM, a stochastic gradient descent method. The code to reproduce this example is available at \url{https://github.com/DecodEPFL/faultDiag}.

	\subsection{Training and Test Data-Sets}  \label{sec:data_sets}
	
	The training and test data-sets are produced by applying various system inputs. These inputs consist of combinations of sinusoidal signals at different frequency ranges to encompass a broad spectrum of input variations. The system input, expressed as $u = [u_1 , u_2]^\top$, is given by
	\begin{equation*} 
		\begin{aligned}
			u_r &= \frac{\max {\tilde \alpha_l}}{\sum_{l=1}^{\tilde n} \tilde \alpha_l} \sum_{l=1}^{\tilde n} \tilde \alpha_l \sin \left(\tilde \omega_l k+\tilde \phi_l\right),\quad r =1,2,  \\
		\end{aligned}
	\end{equation*} 
	for  $k \in\{1,2, \ldots, k_f\}$, where the parameters 
	$$\left(\tilde \alpha_l\right)_{l=1}^{\tilde n}, \left(\tilde \omega_l\right)_{l=1}^{\tilde n}, \left(\tilde \phi_l\right)_{l=1}^{\tilde n},$$ 
	and ${\tilde n}$ are randomly drawn from uniform distributions in the intervals $[0.01,0.1]$ m, $[0.6 \pi,3 \pi]$ rad/s, $[0,0.94\pi]$ rad, and $[2,10]$, respectively. 
	
	The synthetic sensor fault signals $f = [f_1 , f_2, f_3, f_4]^\top$ can be written as  
\begin{equation*} 
	\begin{aligned}
		f_i &= \frac{\max {\underline \alpha_l}}{\sum_{l=1}^{\tilde n} \underline \alpha_l} \sum_{l=1}^{\tilde n} \underline \alpha_l \sin \left(\underline  \omega_l k+\underline \phi_l\right), \quad i =1,2,3,4,  \\
	\end{aligned}
\end{equation*} 
 for $k \in\{1,2, \ldots, k_f\}$, where the parameters 
$$\left(\underline \alpha_l\right)_{l=1}^{\tilde n}, \left(\underline \omega_l\right)_{l=1}^{\tilde n}, \left(\underline \phi_l\right)_{l=1}^{\tilde n},$$ 
are randomly drawn from uniform distributions in the intervals $[0.01,0.1]$ m, $[0.6 \pi,5 \pi]$ rad/s, and $[0,0.94\pi]$ rad, respectively. Note that other fault signals, depending on the faults that actual sensors can experience, can be synthesized at no additional cost. 
	
	To simplify the training process, we have only considered faults in the first and second sensors. We have trained the model using five sets of data for the healthy case, five for each individual faulty sensor, and five for each of the simultaneous faulty sensors (i.e., simultaneous fault in the first and second sensors). This results in in 20 sets of training data (i.e., $s_f = 20$ in \eqref{eq:cost}), which are drawn from the previously described system input and fault distributions. 
	 
	 To evaluate the results, we present one independent test dataset drawn from the same distributions. However, the filter's response for additional test data can be found in the link provided above to the code. Since the studied system has continuous-time dynamics, we have discretized it using the fourth-order Runge-Kutta (RK4) method at 100 Hz, and it is simulated for 20 seconds. For training the fault detection filters, however, the data are downsampled to 4 Hz (i.e., $k_f = 80$ in \eqref{eq:cost}). 
	
	\begin{figure}[t!]
		\centering
		\smallskip
		\includegraphics[width=1\linewidth,keepaspectratio]{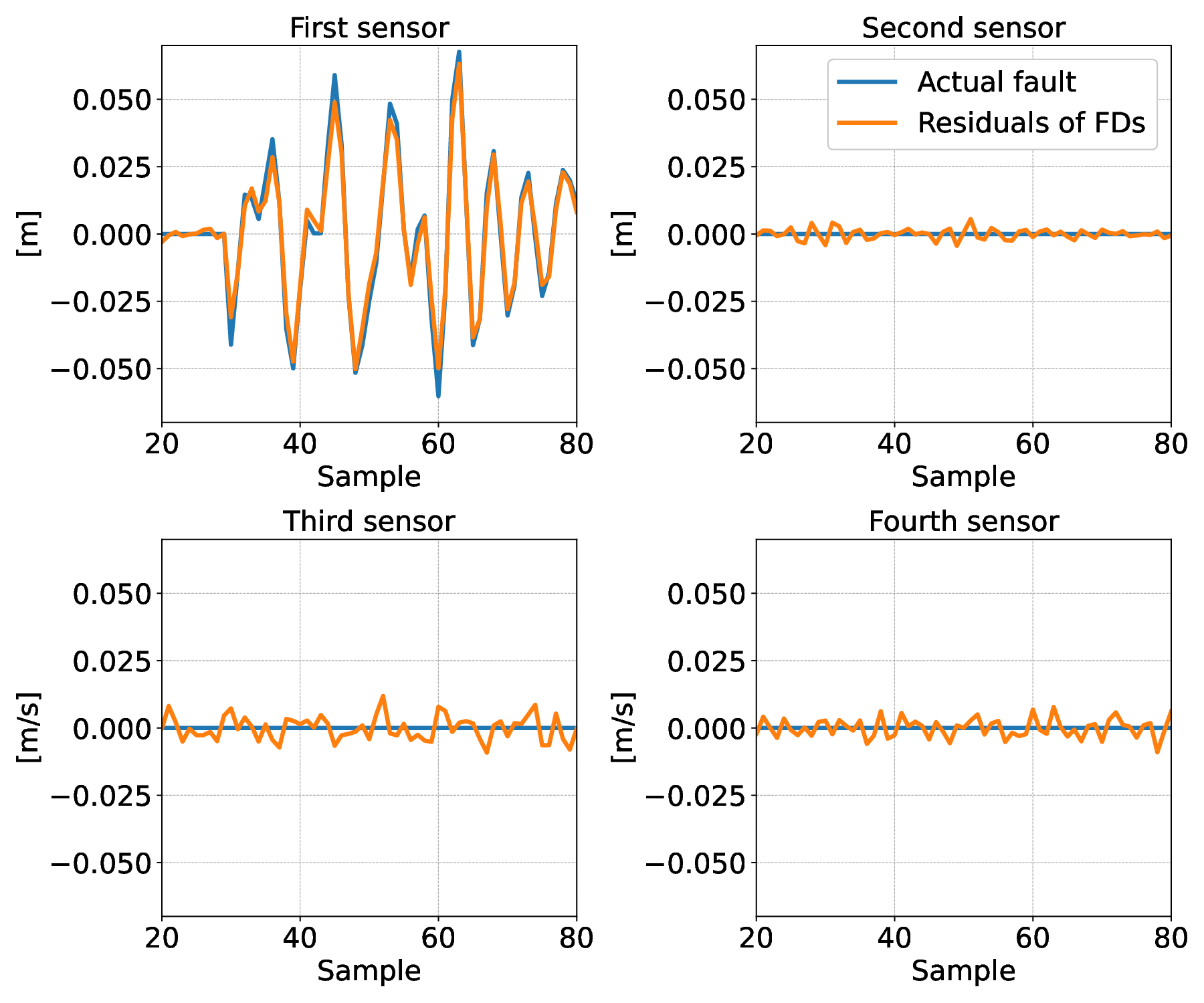}
		\caption{Residual signals of fault detectors for a test scenario with fault in the first sensor.}
		\label{fig:residuals}
	\end{figure}

	\subsection{Performance of Fault Detector Filters}  \label{sec:model_accuraty_rpm}
	
		The residual signals (i.e., the outputs of the FD filters) for all four FD filters, corresponding to a test case where a fault occurs in the first sensor, are depicted in Figure~\ref{fig:residuals}. Each plot corresponds to a filter designed for a specific sensor, as indicated in the titles. As shown, after the fault occurs in the first sensor, its corresponding residual exhibits a clear signal, allowing for fault detection despite the small magnitude of the fault. Furthermore, since the filters are trained to follow the fault signal, the residual generators can serve as fault estimators due to their high accuracy. However, fault estimation is not the focus of this work.

        We have also evaluated the performance and reliability of the designed fault detectors for fault occurrence in the first sensor, the second sensor, and simultaneous faults in both the first and second sensors over the entire set of 1000 test data sets. To assess performance, the mean Root Mean Square Error (RMSE) over the 1000 data sets is computed for different mentioned scenarios, see Table~\ref{tab:results}. Note that in the table, a low mean RMSE for the filter corresponding to the faulty sensor indicates high accuracy in fault signal estimation. On the other hand, lower RMSE values for other filters suggest insensitivity to fault occurrence in the other sensors. Overall, we aim for a low mean RMSE. As can be seen, the error is small, representing the good performance and reliability of the fault detector filters.

\begin{table}[t!] 
\centering
\caption{Mean RMSE values across 1000 test data sets for different fault scenarios.}
\begin{tabular}{|c|c|c|c|c|}
\hline
\begin{tabular}[c]{@{}c@{}}Fault \\ scenario\end{tabular}  & \begin{tabular}[c]{@{}c@{}}First\\ detector\end{tabular} & \begin{tabular}[c]{@{}c@{}}Second\\  detector\end{tabular} & \begin{tabular}[c]{@{}c@{}}Third\\ detector\end{tabular} & \begin{tabular}[c]{@{}c@{}}Fourth\\ detector\end{tabular} \\
\hline
Sensor 1 & 0.0080 & 0.0080 & 0.0047 & 0.0041 \\
Sensor 2 & 0.0072 & 0.0089 & 0.0047 & 0.0041 \\
Sensors 1 \& 2 & 0.0080 & 0.0089 & 0.0047 & 0.0041 \\
\hline
\end{tabular}
\label{tab:results}
\end{table}

    \section{Conclusion} \label{sec:conclusion}
This work has presented a model-free Fault Detection and Isolation (FDI) methodology for discrete-time nonlinear systems, capable of diagnosing single, multiple, and simultaneous sensor faults. Leveraging historical healthy input-output data, the proposed approach has utilized a bank of Robust Recurrent Neural Networks (R-RENs), where each filter is designed to be specifically sensitive to faults in a particular sensor while remaining robust to faults in others. A key advantage of this method is its performance guarantees in terms of the convergence property (providing robustness) and the sensitivity (and insensitivity) to faults. These performance guarantees are inherently imposed by the filter structure.This allows for efficient learning using standard unconstrained optimization techniques. Numerical simulations on a roll-plane model of a vehicle have demonstrated the effectiveness of the proposed methodology for sensor fault detection and isolation. Future work could include providing a direct parameterization for R-REN fault detectors to ensure a sufficient condition for fault sensitivity, as discussed after Problem~\ref{prob:main_prob}. Moreover, it may include the extension of this framework to address more complex fault scenarios, including process and actuator faults.
	
	\appendix
	\section{Appendices}
	\subsection{Direct Parameterization of R-aRENs}   \label{ap:direct_parameteriznatio}
	
	Here, we present the direct parameterization of a R-aREN which can be applied to all the FD filters for the sensors. For the sake of notational simplicity, we omit the subscript $i$ that indicates each FD filter.  
	
	Each FD filter in \eqref{eq:REN_filter} can be expanded as follows:
	\begin{equation} \label{eqapp:ren}
		\left[\begin{array}{c}
			z^+ \\
			v \\
			r
		\end{array}\right]=\overbrace{\left[\begin{array}{c|cc}
				A & B_1 & B_2 \\
				\hline C_1 & D_{11} & D_{12} \\
				C_2 & D_{21} & D_{22}
			\end{array}\right]}^W\left[\begin{array}{l}
			z \\
			w \\
			\bar{u}
		\end{array}\right]+\overbrace{\left[\begin{array}{c}
				\eta_z \\
				\eta_v \\
				\eta_r
			\end{array}\right]}^\eta,
	\end{equation}
	where for a robust acyclic REN, the matrix $D_{11}$ is strictly lower triangular, as imposed by the direct parameterization described next. The above equation can rewritten in the following implicit representation: 
	\begin{equation} \label{eqapp:ren_cvx}
	\left[\begin{array}{c}
		E z^+ \\
		\Lambda v \\
		r
	\end{array}\right]=\overbrace{\left[\begin{array}{ccc}
			F & \mathcal{B}_1 & \mathcal{B}_2 \\
			\mathcal{C}_1 & \mathcal{D}_{11} & \mathcal{D}_{12} \\
			C_2 & D_{21} & D_{22}
		\end{array}\right]}^{\widetilde{W}}\left[\begin{array}{c}
		z \\
		w \\
		\bar{u}
	\end{array}\right]+\tilde{\eta},
	\end{equation}
	where $E$ is an invertible matrix and $\Lambda$ is a positive definite diagonal matrix. Then, the following variables can be freely parameterized in terms of their elements: $\mathcal{B}_2 \in \mathbb{R}^{n_z \times (l+m)}$, $C_2 \in \mathbb{R}^{1 \times n_z}, \mathcal{D}_{12} \in \mathbb{R}^{n_v \times (l+m)}, D_{21} \in \mathbb{R}^{1 \times n_v}, \tilde{\eta} \in \mathbb{R}^{(n_z+n_v+1)}$. The remaining parameters (i.e., $F, \mathcal{B}_1, \mathcal{C}_1, \mathcal{D}_{11}, D_{22}$) should satisfy the following algebraic constraints as they affect contractivity and incremental IQC properties of R-aRENs. Note that some additional free parameters are introduced in what follows. 
	
	The direct parameterization of $D_{22}$ is 
	$$
	D_{22} = L_Q^{-1} N L_R,
	$$
	where $R = L_R^\top L_R, Q = - L_Q^\top L_Q$ are given by the Cholesky decomposition with $R, Q$ as defined in \eqref{eq:RQ} (for $i$-th filter), and 
	$$
	\begin{aligned}
		N &=\left[\begin{array}{ll}
			(I+M)^{-1}(I-M) & -2(I+M)^{-1} Z_3^{\top}
		\end{array}\right],\\
		M &=X_3^{\top} X_3+Y_3-Y_3^{\top}+\epsilon I
	\end{aligned}
	$$
	with $X_3, Y_3$ and $Z_3 \in \mathbb{R}^{(l+m-1)}$ as free variables.
	Next, we construct matrix $H$ as
	\begin{equation}
		\begin{aligned}
			H &= \left[\begin{array}{lll}
				H_{11} & H_{12} & H_{13} \\
				H_{21} & H_{22} & H_{23} \\
				H_{31} & H_{32} & H_{33}
			\end{array}\right] \\
			&= X^{\top} X+\epsilon I+ {\left[\begin{array}{c}
					\mathcal{C}_2^{\top} \\
					\mathcal{D}_{21}^{\top} \\
					\mathcal{B}_2
				\end{array}\right] \mathcal{R}^{-1}\left[\begin{array}{c}
					\mathcal{C}_2^{\top} \\
					\mathcal{D}_{21}^{\top} \\
					\mathcal{B}_2
				\end{array}\right]^{\top}} \\
			&\quad{-\left[\begin{array}{c}
					C_2^{\top} \\
					D_{21}^{\top} \\
					0
				\end{array}\right] Q\left[\begin{array}{c}
					C_2^{\top} \\
					D_{21}^{\top} \\
					0
				\end{array}\right]^{\top} \succ 0, }
		\end{aligned}
	\end{equation}
	where $X \in \mathbb{R}^{(2 n_z+n_v) \times(2 n_z+n_v)}$ is a free variable, $\epsilon$ is a small positive scalar and 
	$$
	\begin{aligned}
		\mathcal{C}_2 &=D_{22}^{\top} Q  C_2, \qquad
		\mathcal{D}_{21} =D_{22}^{\top} Q D_{21}-\mathcal{D}_{12}^{\top}, \\ 
		\mathcal{R}&:=R+D_{22}^{\top} Q D_{22} \succ 0.
	\end{aligned}
	$$
	Then, we partition $H_{22}$ into its diagonal and strictly upper/lower triangular components as
	$$
	H_{22}=\Phi-L-L^{\top},
	$$
	where $\Phi$ is a diagonal matrix and $L$ is a strictly lower triangular matrix. We have: 
	$$
	\Lambda=\frac{1}{2} \Phi, \quad \mathcal{D}_{11}=L.
	$$
	Finally, the rest of the R-aREN variables are specified as follows:
	$$
	\begin{aligned}
		F&=H_{31}, \quad \mathcal{B}_1=H_{32}, \quad \mathcal{C}_1=-H_{21}, \\
		E &=\frac{1}{2}\left(H_{11}+\frac{1}{{\bar{\alpha}}^2} H_{33}+Y_1-Y_1^{\top}\right), \\
	\end{aligned}
	$$
	where $\bar{\alpha} \in(0,1]$ is a given upper bound of the contraction rate of the R-aREN and $Y_1 \in \mathbb{R}^{n_z \times n_z}$ is a free variable. 

	Finally, all the new learnable parameters can be collected in $\theta_d :=\{	\mathcal{B}_2, C_2 , \mathcal{D}_{12}, D_{21}, \tilde{\eta}, X_3, Y_3, Z_3, X, Y_1, \epsilon\}$, which, given the above relations, can be transformed into the parameters of \eqref{eqapp:ren_cvx}. Then, those parameters can easily be mapped to the original parameters of R-RENs, $\theta$, by multiplying the first and second rows of \eqref{eqapp:ren_cvx} by $E^{-1}$ and $\Lambda^{-1}$, respectively.
	
	
	\subsection{Proof of Theorem \ref{theorem}}  \label{ap:thm1_proof}
	Firstly, note that an R-aREN-based FD filter is well-posed if a unique solution is guaranteed. It is shown in \cite[Sec. III.B]{revay2023recurrent} that if the direct parameterization in Appendix~\ref{ap:direct_parameteriznatio} is imposed on the filters \eqref{eq:REN_filter}, then the R-aREN-based filters are well-posed. 
	
	Next, we prove the convergence property for the FD filters \eqref{eq:REN_filter} as R-aRENs. Then, we establish the necessary condition for the sensitivity to faults and subsequently demonstrate the sufficient condition for the insensitivity to other faults.
	
	\textbf{1. Convergence}: Let us first introduce the following lemma, which is used to ensure the convergence property for the FD filters \eqref{eq:REN_filter}.
	
	\begin{lem} \cite[Thm. 3]{jungers2024discrete} \label{lem}
		Consider the general structure of FD filters \eqref{eq:filter_structure}. If there exist a symmetric positive definite matrix $P_i \in \mathbb{R}^{n_z \times n_z},  i = 1,2,\ldots,m$, and a constant $\rho_i$ such that $0<\rho_i<1$ and
		\begin{equation} \label{eqapp:conv_lip}
			\begin{aligned}
			&\left|a_i\left(z^{a}_i, \bar{u}(k)\right)-a_i\left(z^{b}_i, \bar{u}(k)\right)\right|_{P_i} \leq \rho_i\left|z_i^{a}-z_i^{b}\right|_{P_i},\\
			& i = 1,2,\ldots,m,
			\end{aligned}
		\end{equation}
		for all $z^{a}_i, z^{b}_i \in \mathbb{R}^{n_z}$ and any bounded input $\bar{u}(k) \in \mathcal{\bar U}$, and
		\begin{equation} \label{eqapp:conv_bound}
			\left|a_i\left(0, \bar{u}(k)\right)\right|_{P_i}=G_i<+\infty, \quad i = 1,2,\ldots,m,
		\end{equation}
		then, the FD filters \eqref{eq:filter_structure} are globally exponentially convergent for the class of inputs $\mathcal{\bar U}$ that is defined as the set of inputs $\bar{u}$ verifying \eqref{eqapp:conv_bound}. Moreover, the steady-state solutions $\bar{z}_i(k, \bar u), i =1,\ldots,m$, satisfy
		$$
		\left|\bar{z}_i(k, \bar u)\right|_{P_i} \leq \frac{G_i}{1-\rho_i}, \quad i = 1,2,\ldots,m.
		$$
	\end{lem}

	In the proof of Theorem 1 in \cite{revay2023recurrent}, it is shown that for R-RENs (a broader class than R-aRENs), the same incremental inequality as \eqref{eqapp:conv_lip} in Lemma \ref{lem} holds. Now, we only need to show that \eqref{eqapp:conv_bound} holds for R-aRENs. For the sake of notation simplicity, similar to Appendix \ref{ap:direct_parameteriznatio}, we omit the subscript $i$ that indicates each R-aREN filter.  
	
	Given the R-aREN dynamics in \eqref{eqapp:ren}, by setting $z = 0$, we obtain:
	\begin{equation} \label{eqapp:z_and_v}
		\begin{aligned}
			a(0, \bar{u}) &= B_1 w + B_2 \bar{u} + \eta_z, \\
			v   &= D_{11} w + D_{12} \bar{u} + \eta_v,\\
            w &= \sigma\left(v\right).
		\end{aligned}
	\end{equation}
	
	From the second relation in \eqref{eqapp:z_and_v}, given that $\bar{u}$ and $\eta_v$ are bounded, and considering the strictly lower triangularity of matrix $D_{11}$, it can be observed that the vector $v$ is bounded. Consequently, by Assumption~\ref{assum:slope_restricted_nl}, the vector $w = \sigma(v)$, derived from a bounded vector $v$, is also bounded. Furthermore, due to the boundedness of $\bar{u}$, $w$, and $\eta_z$, we can infer that $a(0, \bar{u})$ is bounded, satisfying \eqref{eqapp:conv_bound} for the FD filters. Therefore, we can conclude the convergence property of the FD filters \eqref{eq:REN_filter} using Lemma~\ref{lem}.
	
	\textbf{2. Fault Sensitivity Condition}: Since the direct parameterization in Appendix~\ref{ap:direct_parameteriznatio} is imposed on the filters \eqref{eq:REN_filter}, it is shown in \cite{revay2023recurrent} that the following IQC property \cite[Def. 3]{revay2023recurrent} holds for the filters \eqref{eq:REN_filter}, which are of type R-aREN: 
	\begin{equation} \label{eqapp:IQC}
		\begin{aligned}
		&\sum_{k=0}^{k_f}\left[\begin{array}{l}
			\delta{r_i}^{s_{12}} \\
			\delta{\bar{u}}^{s_{12}} \end{array}\right]^{\top}\left[\begin{array}{cc}
			Q_i & S_i^{\top} \\
			S_i & R_i
		\end{array}\right]\left[\begin{array}{l}
			\delta{r_i}^{s_{12}}\\
			\delta{\bar{u}}^{s_{12}} 
		\end{array}\right] \geq-e(c_i, d_i),  \\
	 &\forall k_f, i = 1,2,\ldots,m,
	 		\end{aligned}
	\end{equation}
	for some function $e(c_i, d_i)$ where $e(c_i, c_i) = 0$, the initial states of the FDs, $c_i$ and $d_i$, correspond to scenarios $s_1$ and $s_2$, respectively. Define $\delta{\bar{u}}^{s_{12}} := \bar{u}^{s_2} - \bar{u}^{s_1}$, where $\bar{u}^{s_1}$ and $\bar{u}^{s_2}$ are the inputs of the FDs  corresponding to scenarios $s_1$ and $s_2$, respectively. Partition the vector $\delta{\bar{u}}^{s_{12}}$ as follows:
	$$
	\delta{\bar{u}}^{s_{12}} = \left[\begin{array}{c}
		\delta{{u}}^{s_{12}} \\
		\delta \tilde{y}_1^{s_{12}} \\
		\delta {y}_i^{s_{12}} \\ 
		\delta \tilde{y}_2^{s_{12}} 
	\end{array}\right] 
$$
with
$$
	\delta \tilde{y}_1^{s_{12}}  := \left[\begin{array}{c}
		\delta {y}_1^{s_{12}} \\ 
		\delta {y}_2^{s_{12}} \\
		\cdots\\  
		\delta {y}_{i-1}^{s_{12}} \\ 
	\end{array}\right], 
	\delta \tilde{y}_2^{s_{12}}  := \left[\begin{array}{c}
		\delta {y}_{i+1}^{s_{12}} \\ 
		\delta {y}_{i+2}^{s_{12}} \\
		\cdots\\  
		\delta {y}_{m}^{s_{12}} \\ 
	\end{array}\right].
	$$
	From \eqref{eqapp:IQC}, by setting $S_i = 0$ and $Q_i,R_i$ as in \eqref{eq:RQ}, it follows that 
	$$
	\begin{aligned}
	&- q_i \| \delta r_i^{s_{12}} \|_{k_f}^2 + \beta_i \| \delta u^{s_{12}} \|_{k_f}^2  + \gamma_i \| \delta \tilde{y}_1^{s_{12}} \|_{k_f}^2  \\
	&\quad + \beta_i \| \delta y_i^{s_{12}} \|_{k_f}^2   +\gamma_i \| \delta \tilde{y}_2^{s_{12}} \|_{k_f}^2 \geq  - e(c_i, d_i).
	\end{aligned}
	$$
	Let us define 	
	$$
	\delta \tilde{y}^{s_{12}}  := \left[\begin{array}{c}
		\delta \tilde{y}_1^{s_{12}} \\ 
		\delta \tilde{y}_2^{s_{12}} \\
	\end{array}\right],  
	$$
	then, the above inequality can be written as 
	\begin{equation} \label{eqapp:beta_gamma}
			\begin{aligned}
		&- q_i \| \delta r_i^{s_{12}} \|_{k_f}^2 + \beta_i \| \delta u^{s_{12}} \|_{k_f}^2  + \beta_i \| \delta y_i^{s_{12}} \|_{k_f}^2   \\
		&\quad+\gamma_i \| \delta \tilde{y}^{s_{12}} \|_{k_f}^2 \geq  - e(c_i, d_i).
			\end{aligned}
	\end{equation}
	Now, consider the healthy case (i.e., where $f$ is zero in \eqref{eq:sys}) as scenario $s_1$ and the occurrence of a fault in the $i$-th sensor as scenario $s_2$. In these two scenarios, the input-output data are identical except for the $i$-th entry of the sensor measurement, $y_i$ (due to the introduction of the fault $f_i$ in the $i$-th sensor). Therefore, the only non-zero incremental term related to the system input-output is the third term in the left-hand side of \eqref{eqapp:beta_gamma}. Moreover, the term on the right-hand side of \eqref{eqapp:beta_gamma} is zero because we have the same initial conditions $c_i = d_i$ for these two scenarios. Therefore, we can conclude that the condition \eqref{eq:fault_sensitivity} holds.
	
	\textbf{3. Other Faults Insensitivity Condition}: Similar to the above results, the incremental inequality \eqref{eqapp:beta_gamma} can be applied between the healthy scenario $s_1$ and a fault occurrence in any sensor except the $i$-th sensor, denoted as scenario $s_3$. In these scenarios, the input-output data are identical except for certain entries of sensor measurements $y_j$ where $j \neq i$ (caused by the fault vector $f = \tilde{f}$ with at least one non-zero entry at $j \neq i$ and zero entry at $i$). Therefore, the only non-zero incremental term related to the system input-output is the fourth term on the left-hand side of \eqref{eqapp:beta_gamma}. Additionally, similar to the previous result, the term on the right-hand side of \eqref{eqapp:beta_gamma} is zero because we have the same initial conditions $c_i = d_i$ for these two scenarios. Consequently, we can conclude that the condition \eqref{eq:faults_insensitivity} holds.
	
	In conclusion, all three properties of the FD filters of the type of R-aRENs in \eqref{eq:REN_filter} have been proven.
	
	\subsection{Roll Plane System Description}  \label{ap:rpm_app}
	The continues equations of motion for the roll plane system are given by \cite{kessels2023real}
	$$M \ddot{\bar q}+f_K(\bar q)+f_C(\dot{\bar q})=f_U(u),$$
	where $\bar q:= [\bar q_1, \bar q_2, \bar q_3, \bar q_4]^\top$ (see Figure~\ref{fig:rpm_schematic}), the mass matrix $M$, the vector containing (linear and nonlinear) spring forces $f_K$, the vector with (linear and nonlinear) damper forces $f_C$, and vector with forces due to the system input $f_U$ are given by
	$$
	\begin{aligned}
		M & =\left[\begin{array}{cccc}
			\frac{\tilde{m}}{2} & \frac{\tilde{m}}{2} & 0 & 0 \\
			-\frac{\tilde{I}}{\tilde{L}} & \frac{\tilde{I}}{\tilde{L}} & 0 & 0 \\
			0 & 0 & \tilde{m}_{t_1} & 0 \\
			0 & 0 & 0 & \tilde{m}_{t_2}
		\end{array}\right], \\
			f_K(\bar q) & =\left[\begin{array}{c}
		F_{K_1}\left(\bar q_1-\bar q_3\right)+F_{K_2}\left(\bar q_2-\bar q_4\right) \\
		\frac{\tilde{L}}{2}\left(F_{K_2}\left(\bar q_2-\bar q_4\right)-F_{K_1}\left(\bar q_1-\bar q_3\right)\right) \\
		F_{K_1}\left(\bar q_3-\bar q_1\right)+\tilde{k}_{t_1} \bar q_3 \\
		F_{K_2}\left(\bar q_4-\bar q_2\right)+\tilde{k}_{t_2} \bar q_4
	\end{array}\right], \\
		\end{aligned}
	$$
	$$
		\begin{aligned}
		f_C(\bar q)&  =\left[\begin{array}{c}
			F_{C_1}\left(\dot{\bar q}_1-\dot{\bar q}_3\right)+F_{C_2}\left(\dot{\bar q}_2-\dot{\bar q}_4\right) \\
			\frac{\tilde{L}}{2}\left(F_{C_1}\left(\dot{\bar q}_3-\dot{\bar q}_1\right)-F_{C_2}\left(\dot{\bar q}_4-\dot{\bar q}_2\right)\right) \\
			F_{C_1}\left(\dot{\bar q}_3-\dot{\bar q}_1\right) \\
			F_{C_2}\left(\dot{\bar q}_4-\dot{\bar q}_2\right)
		\end{array}\right], \\
		f_U(u)& =\left[\begin{array}{c}
			0 \\
			0 \\
			\tilde{k}_{t_1} u_1(t) \\
			\tilde{k}_{t_2} u_2(t)
		\end{array}\right],
	\end{aligned}
	$$
	where nonlinear stiffness and damper terms are as follows:
	$$
	\begin{aligned}
		F_{K_s}(\bar q) & =\tilde{k}_s \bar q+\tilde{k}_{s_n} \bar q^3, \\
		F_{C_s}(\dot{\bar q}) & = = \tilde{c}_s \dot{\bar q} + \tilde{c}_{s,n}(0.2 \tanh (10 \dot{\bar q})), \quad s = 1, 2.
	\end{aligned}
	$$
The parameter values are listed in Table \ref{tab:rpm_parameters}.
\begin{table}[tb!] 
	\caption{Parameter values of roll plane system.}
	\centering
	\begin{tabular}{ccc}
		\hline
		Parameter                                                           & Value  & Unit   \\ \hline
		$\tilde{m}$ &               $580$                                                                      &   $\mathrm{~kg}$                                                                                                                      \\ [1.5mm]
		$\tilde{m}_{t_1}, \tilde{m}_{t_2}$ &                          $36.26$                                                            &   $\mathrm{~kg}$                                                                                                                                                \\ [1.5mm]
		$\tilde{I}$ &                          $63.3316$                                                            &   $\mathrm{~kg} \cdot \mathrm{m}^2$                                                                                                                                               \\ [1.5mm]
		$\tilde{L}$ &                          $1.524$                                                            &   $\mathrm{~m}$                                                                                                                                                \\ [1.5mm]
		$\tilde{c}_1, \tilde{c}_2$ &                          $710.70$                                                            &   $\frac{\mathrm{N} \cdot \mathrm{s}}{\mathrm{m}}$                                                                                                                                                \\ [1.5mm] 
		$\tilde{c}_{1,n}, \tilde{c}_{2,n}$ &                  $0.71$                                                            &   $\frac{\mathrm{N} \cdot \mathrm{s}}{\mathrm{m}}$                                                                                                                                                \\ [1.5mm] 
		$\tilde{k}_1, \tilde{k}_2$ &                          $19357.2$                                                            &   $\frac{\mathrm{N}}{\mathrm{m}}$    \\ [1.5mm]
		$\tilde{k}_{t_1}, \tilde{k}_{t_2}$ &                          $96319.76$                                                            &   $\frac{\mathrm{N}}{\mathrm{m}}$                                                                                                                                        \\ [1.5mm]
		$\tilde{k}_{1_n}, \tilde{k}_{2_n}$ &                          $15000$                                                            &   $\frac{\mathrm{N}}{\mathrm{m}^3}$                                                                                                                                         \\ \hline
	\end{tabular} \label{tab:rpm_parameters}
\end{table}

	\section*{acknowledgment}
	This publication is part of the project Digital Twin project
	4.3 with project number P18-03 of the research programme
	Perspectief which is (mainly) financed by the Dutch Research
	Council (NWO).

	\bibliographystyle{IEEEtran}
	\bibliography{IEEEabrv,refs}

\end{document}